\documentclass[aps,pre,twocolumn,showpacs,showkeys,nofootinbib]{revtex4}

\usepackage{amsmath}
\usepackage{amsfonts}
\usepackage{amssymb}
\usepackage[english]{babel}
\usepackage{bm}    %% For bold math symbols
\usepackage{epsf}

\begin{document}

\title{Systematic errors due to linear congruential
       random-number generators \\
       with the Swendsen--Wang algorithm:
       A warning}

\author{Giovanni Ossola}
\email{giovanni.ossola@physics.nyu.edu}
\author{Alan D.~Sokal}
\email{sokal@nyu.edu}
\affiliation{Department of Physics, New York University,
  4 Washington Place, New York, NY 10003, USA}

\date{8 March 2004}

\begin{abstract}
We show that linear congruential pseudo-random-number generators
can cause systematic errors in Monte Carlo simulations using
the Swendsen--Wang algorithm,
if the lattice size is a multiple of a very large power of 2
and one random number is used per bond.
These systematic errors arise from correlations within a single
bond-update half-sweep.
The errors can be eliminated (or at least radically reduced)
by updating the bonds in a random order or in an aperiodic manner.
It also helps to use a generator of large modulus (e.g.\ 60 or more bits).
\end{abstract}



\maketitle

\newtheorem{defin}{Definition}[section]
\newtheorem{definition}[defin]{Definition}
\newtheorem{prop}[defin]{Proposition}
\newtheorem{proposition}[defin]{Proposition}
\newtheorem{lem}[defin]{Lemma}
\newtheorem{lemma}[defin]{Lemma}
\newtheorem{guess}[defin]{Conjecture}
\newtheorem{ques}[defin]{Question}
\newtheorem{question}[defin]{Question}
\newtheorem{prob}[defin]{Problem}
\newtheorem{problem}[defin]{Problem}
\newtheorem{thm}[defin]{Theorem}
\newtheorem{theorem}[defin]{Theorem}
\newtheorem{cor}[defin]{Corollary}
\newtheorem{corollary}[defin]{Corollary}
\newtheorem{conj}[defin]{Conjecture}
\newtheorem{conjecture}[defin]{Conjecture}

\newtheorem{pro}{Problem}
\newtheorem{clm}{Claim}
\newtheorem{con}{Conjecture}

%
% For examples (numbered by sections, separately from theorems etc)
% Note that the \label does not include the section number;
%   that has to be put in by hand in the \ref.
%
\newcounter{example}[section]
\newenvironment{example}%
{\refstepcounter{example}
 \bigskip\par\noindent{\bf Example \thesection.\arabic{example}.}\quad
}%
%{\hbox{\kern1pt\vrule height6pt width4pt depth1pt\kern1pt}}
%{\hbox{\hskip 6pt\vrule width6pt height7pt depth1pt \hskip1pt}}
{\quad $\Box$}
\def\bexam{\begin{example}}
\def\eexam{\end{example}}

\renewcommand{\theenumi}{\alph{enumi}}
\renewcommand{\labelenumi}{(\theenumi)}
\def\prf{\par\noindent{\bf Proof.\enspace}\rm}
\def\rmk{\par\medskip\noindent{\bf Remark.\enspace}\rm}

\newcommand{\be}{\begin{equation}}
\newcommand{\ee}{\end{equation}}
\newcommand{\bmath}{\begin{displaymath}}
\newcommand{\emath}{\end{displaymath}}
\newcommand{\<}{\langle}
\renewcommand{\>}{\rangle}
\newcommand{\widebar}{\overline}
\def\reff#1{(\protect\ref{#1})}
\def\spose#1{\hbox to 0pt{#1\hss}}
\def\ltapprox{\mathrel{\spose{\lower 3pt\hbox{$\mathchar"218$}}
 \raise 2.0pt\hbox{$\mathchar"13C$}}}
\def\gtapprox{\mathrel{\spose{\lower 3pt\hbox{$\mathchar"218$}}
 \raise 2.0pt\hbox{$\mathchar"13E$}}}
\def\textprime{${}^\prime$}
\def\proof{\par\medskip\noindent{\sc Proof.\ }}
\newcommand{\qed}{\quad $\Box$ \medskip \medskip}
\def\proofof#1{\bigskip\noindent{\sc Proof of #1.\ }}
\def\half{ {1 \over 2} }
\def\third{ {1 \over 3} }
\def\twothird{ {2 \over 3} }
\def\smfrac#1#2{\textstyle{#1\over #2}}
\def\smhalf{ \smfrac{1}{2} }
\newcommand{\real}{\mathop{\rm Re}\nolimits}
\renewcommand{\Re}{\mathop{\rm Re}\nolimits}
\newcommand{\imag}{\mathop{\rm Im}\nolimits}
\renewcommand{\Im}{\mathop{\rm Im}\nolimits}
\newcommand{\sgn}{\mathop{\rm sgn}\nolimits}
\newcommand{\var}{\mathop{\rm var}\nolimits}
\newcommand{\cov}{\mathop{\rm cov}\nolimits}
\def\hboxscript#1{ {\hbox{\scriptsize\em #1}} }

\newcommand{\restrict}{\upharpoonright}
\renewcommand{\emptyset}{\varnothing}

\def\Z{{\mathbb Z}}
\def\ZZ{{\mathbb Z}}
\def\R{{\mathbb R}}
\def\C{{\mathbb C}}
\def\CC{{\mathbb C}}
\def\N{{\mathbb N}}
\def\NN{{\mathbb N}}
\def\Q{{\mathbb Q}}

\newcommand{\scra}{{\mathcal{A}}}
\newcommand{\scrb}{{\mathcal{B}}}
\newcommand{\scrc}{{\mathcal{C}}}
\newcommand{\scrf}{{\mathcal{F}}}
\newcommand{\scrg}{{\mathcal{G}}}
\newcommand{\scrh}{{\mathcal{H}}}
\newcommand{\scrl}{{\mathcal{L}}}
\newcommand{\scro}{{\mathcal{O}}}
\newcommand{\scrp}{{\mathcal{P}}}
\newcommand{\scrr}{{\mathcal{R}}}
\newcommand{\scrs}{{\mathcal{S}}}
\newcommand{\scrt}{{\mathcal{T}}}
\newcommand{\scrv}{{\mathcal{V}}}
\newcommand{\scrw}{{\mathcal{W}}}
\newcommand{\scrz}{{\mathcal{Z}}}
\newcommand{\scrbt}{{\mathcal{BT}}}
\newcommand{\scrbf}{{\mathcal{BF}}}

\newcommand{\bgamma}{\boldsymbol{\gamma}}
\newcommand{\bsigma}{\boldsymbol{\sigma}}
\renewcommand{\pmod}[1]{\;({\rm mod}\:#1)}

% Array for subscripts

\newenvironment{sarray}{
	  \textfont0=\scriptfont0
	  \scriptfont0=\scriptscriptfont0
	  \textfont1=\scriptfont1
	  \scriptfont1=\scriptscriptfont1
	  \textfont2=\scriptfont2
	  \scriptfont2=\scriptscriptfont2
	  \textfont3=\scriptfont3
	  \scriptfont3=\scriptscriptfont3
	\renewcommand{\arraystretch}{0.7}
	\begin{array}{l}}{\end{array}}

\newenvironment{scarray}{
	  \textfont0=\scriptfont0
	  \scriptfont0=\scriptscriptfont0
	  \textfont1=\scriptfont1
	  \scriptfont1=\scriptscriptfont1
	  \textfont2=\scriptfont2
	  \scriptfont2=\scriptscriptfont2
	  \textfont3=\scriptfont3
	  \scriptfont3=\scriptscriptfont3
	\renewcommand{\arraystretch}{0.7}
	\begin{array}{c}}{\end{array}}

\section{Introduction} \label{sec1}

It has been known for about two decades that
linear congruential pseudo-random-number generators \cite{Knuth_81}
suffer from strong long-range correlations
\cite{Kalle_84,Filk_85,Percus_88,DeMatteis_88,Eichenauer_89}:
for instance, generators with modulus $m = 2^\beta$
have strong correlations at lags that are multiples of $2^k$
whenever the ratio $k/\beta$ is large enough
\cite{Kalle_84,Filk_85,Percus_88}.
Furthermore, these long-range correlations are known to give rise
to systematic errors in Monte Carlo simulations employing
local (e.g.\ Metropolis or heat bath) updates
whenever the lattice sites are updated in a fixed order
and the number of random numbers used per sweep
is a multiple of a large-enough power of~2:
this happens because one is using strongly correlated random numbers
to update the same lattice sites in successive sweeps
(within roughly one autocorrelation time)
\cite{Kalle_84,Filk_85,Montvay_87,Damgaard_88,Berg_89,Baig_95}.
On the other hand, these systematic errors can be eliminated
by the simple expedient of throwing away one random number
at the end of each lattice sweep
\cite{Kalle_84,Filk_85,Montvay_87,Damgaard_88,Berg_89,Baig_95}.

It has generally been thought that non-local algorithms
such as the Swendsen--Wang algorithm \cite{Swendsen_87}
and Wolff's single-cluster variant \cite{Wolff_89}
would be immune to these particular defects of
linear congruential generators,
inasmuch as they employ random numbers in a highly aperiodic way
both in ``space'' and in ``time''.
We were therefore astonished to find,
in our Swendsen--Wang simulation of the three-dimensional Ising model
\cite{Ossola-Sokal},
large systematic errors on the $128^3$ and $256^3$ lattices
that we eventually traced (after much wringing of hands)
precisely to long-range correlations in the random-number generator.

Recall that one iteration of the Swendsen--Wang (SW) algorithm
consists of two steps:
first one updates the bond occupation variables
at a fixed configuration of the Ising spin variables;
then one computes the connected clusters associated to the
bond configuration and updates the Ising spin variables
by choosing a new spin value independently for each cluster.
The second (spin-update) half of the SW algorithm indeed uses random numbers
in a thoroughly aperiodic way,
because the cluster sizes and shapes are random.
But the first (bond-update) half uses random numbers
in a highly structured way:
typically one sweeps the bonds of the lattice in some simple fixed order
(e.g.\ lexicographic).
Therefore, if the lattice size is very large,
the effects of the long-range correlations of the 
random-number generator can be observed
{\em within a single half-sweep}\/:
the random numbers used in updating the bonds of
one part of the lattice will be strongly correlated
with those used elsewhere in the lattice.
One may expect this correlation to cause systematic errors particularly if
(a) the lattice size is commensurate with the lag giving rise
to long-range correlations (e.g.\ a power of 2),
and (b) the system's correlation length is large enough so that the
long-range correlations of the random-number generator
couple {\em correlated}\/ parts of the lattice.

The purpose of this note is, first of all,
to provide evidence that such systematic errors can indeed occur
and that we have accurately diagnosed their origin;
and secondly, to show how the implementation of the Swendsen--Wang algorithm
can be modified so as to eliminate (or at least radically reduce)
these systematic errors.
A more detailed account will be published elsewhere \cite{systematic_full}.

% The plan of this paper is as follows:
% In Section~\ref{sec2} we review the number theory underlying
% the long-range correlations in linear congruential generators;
% along the way we make some small new contributions.
% In Section~\ref{sec3} we report our data from the
% three-dimensional Ising model that demonstrate the existence
% of the systematic errors and give evidence as to their origin.
% In Section~\ref{sec4} we propose a theoretical model
% for the systematic errors and compare it with our Monte Carlo data.
% Finally, in Section~\ref{sec_discussion} we discuss our results
% in the context of some prior work
% on the defects of random-number generators
% and their effects on Monte Carlo simulations
% \cite{Ferrenberg_92,Grassberger_93a,Grassberger_93b,Selke_93,Vattulainen_94,%
% Coddington_94,Coddington_96}.
% {\bf Follow Coddington and criticize Landau {\em et al.}\/???}

\section{Evidence of systematic errors}

We simulated the nearest-neighbor three-dimensional Ising model
on an $L \times L \times L$ simple-cubic lattice
with periodic boundary conditions,
using the Swendsen--Wang (SW) algorithm \cite{Swendsen_87}.
We studied lattice sizes
$L=4$, 6, 8, 12, 16, 24, 32, 48, 64, 96, 128, 192, 256
and performed between $10^7$ and $10^8$ SW iterations for each lattice size.
We did all our runs at $\beta = 0.22165459$,
which is Bl\"ote {\em et al.}\/'s best estimate of the critical temperature
\cite{Blote_99}
and is very near to the estimates by other workers
\cite{Ballesteros_99,Hasenbusch_99}
(see also the review \cite{Hasenbusch_01}).
We measured a large number of observables,
including the susceptibility $\chi$,
the second-moment correlation length $\xi$,
the energy $E$, and the specific heat $C_H$.

In the first version of our program, the random numbers were supplied
by a linear congruential generator with modulus $m=2^{48}$,
increment $c=1$, and multiplier
$a=31167285$, 10430376854301, 77596615844045 or 181465474592829.
All these multipliers give good results on the spectral test in low dimensions,
compared to other multipliers for the same modulus \cite{Knuth_81,Lecuyer_99}.
We verified that the runs with the four different multipliers
gave results that are consistent within error bars
for all the major observables;
after making this verification,
we averaged all the runs for each $L$.

\begin{table}[t]
\begin{center}
\begin{tabular}{|r||l|r|r|}
\hline
   \multicolumn{1}{|c||}{$L$} &  \multicolumn{1}{|c|}{$\xi/L$}  &
   deviation (\%) & deviation ($\sigma$) \\
\hline
     4 &  0.63000(9)   & $-$0.566\%  &  ${\bf -17.48} \bsigma$ \\
     6 &  0.63576(10)  & $-$0.076\%  &  ${\bf -3.24} \bsigma$ \\
     8 &  0.63769(10)  &    0.004\%  &  $0.44 \sigma$ \\
    12 &  0.63909(11)  & $-$0.013\%  &  $-0.83 \sigma$ \\
    16 &  0.63998(8)   &    0.002\%  &  $0.15 \sigma$ \\
    24 &  0.64093(13)  &    0.015\%  &  $0.83 \sigma$ \\
    32 &  0.64122(13)  & $-$0.010\%  &  $-0.51 \sigma$ \\
    48 &  0.64172(10)  & $-$0.007\%  &  $-0.51 \sigma$ \\
    64 &  0.64223(16)  &    0.032\%  &  $1.44 \sigma$ \\
    96 &  0.64219(16)  & $-$0.017\%  &  $-0.76 \sigma$ \\
   128 &  0.62215(25)  & $-$3.159\%  &  ${\bf -78.94} \bsigma$ \\
   192 &  0.64383(38)  &    0.192\%  &  ${\bf 3.23} \bsigma$ \\
   256 &  0.77798(79)  &   21.052\%  &  ${\bf 169.69} \bsigma$ \\
\hline
$\infty$ & 0.64299(8)  & \multicolumn{2}{c}{\quad}    \\
\cline{1-2}
\end{tabular}
\vspace{5mm}
\caption{
   Results of our Swendsen--Wang simulations on the three-dimensional
   Ising model at criticality, using a linear congruential generator
   with modulus $m=2^{48}$.
   Error bar (one standard deviation) is shown in parentheses.
   The row marked $L=\infty$ indicates our best estimate of
   the asymptotic value $x^\star$.
   The last two columns indicate the deviation of each point
   from the fit curve \reff{fit_xstar}, in percent and in standard deviations.
   Points deviating by more than $3\sigma$ are marked in boldface.
}
\label{table_3Dising_1}
\end{center}
\end{table}

The results for the correlation length $\xi$ are reported in
the first two columns of Table~\ref{table_3Dising_1}.
Finite-size-scaling theory predicts that $\xi/L$ should behave
for large $L$ (if indeed we are at the critical temperature) as
\be
   \xi/L  \;=\;  x^\star \,+\, A L^{-\omega} \,+\, \ldots \;,
 \label{def_FSS}
\ee
where $x^\star$ is a universal amplitude ratio
characteristic of the given system with periodic boundary conditions,
$\omega$ is a correction-to-scaling exponent,
$A$ is a nonuniversal correction-to-scaling amplitude,
and the dots indicate higher-order corrections to scaling.
The data in Table~\ref{table_3Dising_1} are qualitatively consistent
with \reff{def_FSS},
{\em except for the points at $L=128$ and $L=256$,
 which show extremely large deviations}\/.

A closer examination of the data in Table~\ref{table_3Dising_1}
reveals that the point at $L=192$ may also exhibit
a small but statistically significant deviation from the fitting curve.
To make all these observations more quantitative,
let us perform a weighted least-squares fit to \reff{def_FSS}
with $\omega = 0.82$ (the best estimate from \cite{Blote_99}),
using all the data with $L_{\rm min} \le L \le 96$
and varying $L_{\rm min}$ while checking the goodness of fit.
A good fit ($\chi^2 = 3.85$, 6 DF, confidence level = 70\%)
can be obtained already with $L_{\rm min} = 8$, yielding
\be
   \xi/L  \;\approx\; 0.64299(8) - 0.02931(79) L^{-0.82}  \;.
 \label{fit_xstar}
\ee
Not surprisingly, the points $L=4$ and $L=6$ show significant deviations
from the fit curve, due to higher-order corrections to scaling.
More surprising are the points $L=128$, which lies roughly 3\%
($\approx 79$ standard deviations) below the fit curve,
and $L=256$, which lies a whopping 21\%
($\approx 170$ standard deviations) above the fit curve.
Obviously something has gone badly wrong!
Finally, the point $L=192$ lies approximately 0.2\%
($\approx 3$ standard deviations) above the fit curve:
this {\em may}\/ indicate the presence
of a small systematic error also for this lattice.

At first we worried whether we had made a programming error
that might lead to incorrect results on large lattices
(e.g.\ due to integer overflow).
We checked the program carefully and were unable to find any such mistakes.
Moreover, the fact that the systematic discrepancy
is much smaller (if it exists at all) at $L=192$ than at $L=128$
suggests that the problem --- whatever its cause ---
does not arise {\em solely}\/ from the lattice being large.

Intrigued by the fact that these large discrepancies might be arising
only at lattice sizes that are large powers of 2
(or perhaps multiples of large powers of 2),
we made shorter runs (between $3 \times 10^4$ and $10^6$ SW iterations)
at many other lattice sizes ---
all multiples of 2 from 4 through 140,
and all multiples of 10 through 250 ---
in order to check whether any other deviant points could be found.
The upshot is that
--- to within the statistical error of these shorter runs,
which ranges from 0.2\% on small lattices
 to 1\% at $L \approx 128$
 to an admittedly rather crude 2.3\% on the largest 
lattices ---
there are no detectable discrepancies
except at $L=128$ and 256.

At $L=128$ and $L=256$ we found discrepancies
not only for the correlation length
but also for the susceptibility, the energy and the specific heat.
It is a curious fact, however, that all the
Fortuin--Kasteleyn identities
\cite[equations (3.20)--(3.23)]{Salas-Sokal_00}
are verified perfectly (to within statistical error).
This contrasts with the systematic errors found by
Damgaard and Heller \cite{Damgaard_88}
in a Metropolis Monte Carlo simulation of the $U(1)$ Higgs model,
where a Ward identity was violated by up to 10 standard deviations,
and those found by Ballesteros and Mart\'{\i}n-Mayor \cite{Ballesteros_98}
in a Wolff single-cluster simulation of the two- and three-dimensional
Ising models, in which Schwinger--Dyson identities were violated by
up to 8 standard deviations.

For several weeks we had no idea what might be causing
the systematic discrepancies at $L=128$ and $L=256$.
(We felt like a detective with a corpse on his hands
 but no suspect and no modus operandi.)
But then we realized, as explained in the Introduction,
that long-range correlations in the random-number generator
could cause undesired correlations within a single bond-update sweep.

\section{Variant simulations}

In order to test whether our proposed explanation
for the systematic errors is the correct one,
we ran variant simulations in which two aspects of the simulation
were systematically altered:
the modulus $m=2^\beta$ of the random-number generator
($\beta=16$, 20, 24, 28, 32, 40, 48, 60, 63, 64),
and the manner in which the random numbers are used within the
bond-update subroutine.
The latter test is essential if we are to prove not only
that the trouble comes from the random-number generator,
but more specifically that it comes from
the way that the random numbers are used
{\em in the bond-update subroutine}\/.

All of the multipliers used here
give good results on the spectral test in low dimensions
compared to other multipliers for the same modulus.\footnote{
   For moduli $2^{32}$, $2^{40}$, $2^{48}$, $2^{60}$, $2^{63}$ and $2^{64}$,
   these multipliers can be found in published tables of multipliers
   that perform well on the spectral test \cite{Knuth_81,Lecuyer_99}.
   We double-checked these computations,
   and performed analogous computations from scratch for the smaller moduli.
}
The purpose of trying random-number generators with less than 48 bits
was to induce systematic errors on small lattices where they could be studied
quantitatively to high precision and compared with those observed
with the 48-bit generator on larger lattices.
The purpose of trying random-number generators with 60/63/64 bits
was, of course, to provide a standard of comparison
in which the systematic error is eliminated or at least radically reduced.
% \begin{itemize}
%    \item {\bf 32-bit:}  Modulus $m=2^{32}$ and multiplier
%        $a=32310901$.
%        The purpose of trying a 32-bit generator was to induce
%        systematic errors on small lattices where they could be studied
%        quantitatively to high precision and compared with those observed
%        with the 48-bit generator on larger lattices.
%    \item {\bf 48-bit:}   Modulus $m=2^{48}$ and multipliers
%        $a=31167285$, 10430376854301, 77596615844045 and 181465474592829.
%    \item {\bf 60/63/64-bit:}
% \begin{itemize}
%    \item[{}] Modulus $m=2^{60}$, multiplier $a=454339144066433781$.
%    \item[{}] Modulus $m=2^{63}$, multiplier $a=9219741426499971445$.
%    \item[{}] Modulus $m=2^{64}$, multiplier $a=3202034522624059733$.
% \end{itemize}
% \end{itemize}

We also tried three variants of the bond-update subroutine:

   {\em Standard:}\/  This is our original program, in which the
       bonds are updated in lexicographic order,
       and one random number is used per bond.

   {\em Aperiodic:}\/
       Here the bonds are again updated in
       lexicographic order, but a random number is used only if
       the two spins are equal.  (If the two spins are unequal,
       the corresponding bond is automatically left unoccupied,
       so no random number is needed.)
       If our explanation of the cause of the systematic errors is correct,
       this strategem should eliminate the systematic errors on
       lattices that are multiples of large powers of 2,
       though it may conceivably shift those systematic errors
       to other lattice sizes
       (namely, those for which the lattice size, multiplied by the
        fraction of nearest-neighbor spins that are equal,
        yields a suitable ``resonance'').

   {\em Shuffle:}\/  The bonds are updated in a random order.\footnote{
         A uniform random permutation of $n$ elements can easily be
         enerated, in a time of order $n$, using $n-1$ random numbers
         \protect\cite[pp.~139--140]{Knuth_81}.
}
      If our explanation of the cause of the systematic errors is correct,
      this strategem should entirely eliminate the systematic errors,
      even with a relatively poor (e.g.\ 32-bit) random-number generator.

      Our first version of the ``shuffle'' subroutine permuted
      the array containing the bond indices.
      Unfortunately, this program ran very slowly ---
      about a factor of 2 slower than the ``standard'' version at $L=16$,
      growing to a factor $\approx 8$ at $L=256$ ---
      probably because the highly nonlocal access to the bond array
      caused a large number of cache misses.
      Our second version permuted instead
      the array of random numbers\footnote{
         More precisely, it permuted a {\sc LOGICAL} array
         containing the results of the comparisons of the random numbers
         against $p = 1 - e^{-2\beta}$.
         This requires only one byte storage per bond,
         rather than 8 bytes for storing the random number itself,
         thereby reducing both memory usage and cache misses during the
         generation of the random permutation.
};
      this is statistically equivalent but allows the bond array
      to be accessed in sequential order.
      This program ran less slowly:  once again
      about a factor of 2 slower than the ``standard'' version at $L=16$,
      but growing only to a factor $\approx 4$ at $L=256$.

% It is worth remarking that we did not initially plan to go to
% all this effort!
% Rather, we started by trying one 32-bit generator and one 60-bit generator;
% shortly afterwards, we realized that we could also implement
% 63-bit and 64-bit generators in our software, and we did so.
% But the results from these initial simulations were puzzling to us:
% the magnitude of the systematic errors
% as a function of modulus (32-bit versus 48-bit) and lattice size ($L$)
% did not seem consistent with any of our ``scaling'' theories.
% So, in an attempt to obtain more refined information upon which to build
% a revised scaling theory, we began to try the other moduli:
% first 24-bit and 40-bit, and then the rest.
% And when we noticed that the results for small moduli seem to depend not only
% on the modulus but also on the specific multiplier used,
% we tried additional multipliers for the same modulus.

The results of all these variant simulations,
carried out on lattice sizes $L=8,16,32,64,96,128,192,256$,
will be reported elsewhere \cite{systematic_full};
here we provide only a brief summary.
We find that the 60/63/64-bit generators give consistent results
(within statistical error) for all three variants of the
bond-update subroutine, confirming our expectation that they exhibit
negligible systematic error on lattices $L \le 256$.\footnote{
   We are continuing runs on the lattice $L=256$ in an effort
   to detect very small systematic errors.
   These results will be reported later \cite{systematic_full}.
}
By contrast, each ``standard'' algorithm with $\le 48$ bits
exhibits detectable systematic errors whenever the lattice size $L$
is a multiple of a sufficiently large power of 2;
how large depends on the modulus.
More precisely, the 16-bit
(resp.\ 20-bit, 24-bit, 28-bit, 32-bit, 40-bit, 48-bit)
standard algorithm
exhibits detectable systematic errors whenever $L$
is a multiple of 8 (resp.\ 8, 8, 16, 32, 64, 128).
In addition, the 48-bit ``standard'' algorithm at $L=192$
shows a discrepancy of almost $3\sigma$,
which {\em may}\/ indicate a systematic error.
No other statistically significant discrepancies are observed.

We conclude that, if one wants to use a linear congruential generator
with the Swendsen--Wang algorithm, the safest approach is to use
a generator of 64 bits (or more) together with the ``shuffle'' bond update.
Unfortunately, the shuffle method is somewhat slow.
A much faster --- and, as far as we can tell, also safe --- method
is to use a 64-bit generator together with the ``aperiodic'' bond update.

Despite the known problems of linear congruential generators
arising from long-range correlations, there are still several advantages
in using them.
First, they are relatively cheap in terms of CPU time,
and are convenient for use in a series of successive runs
because the complete state of the generator
can be saved in a single computer word.
More importantly, they are well understood theoretically,
as regards both short-range \cite{Knuth_81}
and long-range
\cite{Kalle_84,Filk_85,Percus_88,DeMatteis_88,Eichenauer_89,systematic_full}
correlations;
in particular, excellent equidistribution of $t$-tuples
of successive random numbers for small $t$
can be achieved by careful choice of the multiplier.
By contrast, for more exotic random-number generators
(e.g.\ combination generators),
the problems may not be absent, but simply hidden.

A more detailed analysis of these simulations
will be published elsewhere \cite{systematic_full},
along with a discussion of the advantages and disadvantages
of linear congruential versus other types
of pseudo-random-number generators
(see also \cite{Ferrenberg_92,Vattulainen_94,Coddington_94,Ballesteros_98}).

\begin{acknowledgments}
We wish to thank Henk Bl\"ote, Lu\'{\i}s Antonio Fern\'andez,
Werner Kerler, Juan Ruiz-Lorenzo, Lev Shchur and Jian-Sheng Wang
for valuable correspondence and for sharing their unpublished data;
Mulin Ding and Raj Sivanandarajah for efficient assistance
with numerous aspects of the Physics Department computing system;
and Rom\'an Scoccimarro for generously letting us
do some test runs on his DEC Alpha computer.
This research was supported in part by
U.S.\ National Science Foundation grants PHY--0099393
and PHY--0116590. %% This is the MRI computer grant
\end{acknowledgments}


\begin{thebibliography}{99}

\bibitem{Knuth_81}  D.E. Knuth, {\em The Art of Computer Programming}\/,
   vol.~2, 2nd ed. (Addison-Wesley, Reading, Massachusetts, 1981).

\bibitem{Kalle_84}  C. Kalle and S. Wansleben, Computer Phys. Commun.
   {\bf 33}, 343 (1984).

\bibitem{Filk_85}  T. Filk, M. Marcu and K. Fredenhagen,
   Phys. Lett. B {\bf 165}, 125 (1985).

\bibitem{Percus_88}  O.E. Percus and J.K. Percus, J. Comput. Phys. {\bf 77},
   267 (1988).

\bibitem{DeMatteis_88}  A. DeMatteis and S. Pagnutti, Numer. Math. {\bf 53},
   595 (1988).

\bibitem{Eichenauer_89}  J. Eichenauer-Herrmann and H. Grothe,
   Numer. Math. {\bf 56}, 609 (1989).

\bibitem{Montvay_87}  I. Montvay and P. Weisz, Nucl. Phys. B {\bf 290},
   327 (1987), see p.~340.

\bibitem{Damgaard_88}  P.H. Damgaard and U.M. Heller,
   Nucl. Phys. B {\bf 309}, 625 (1988), see p.~639.

\bibitem{Berg_89}  B.A. Berg and A.H. Billoire, Phys. Rev. D {\bf 40}, 550
   (1989), see Appendix B.

\bibitem{Baig_95}  M. Baig, H. Fort, J.B. Kogut and S. Kim,
   Phys. Rev. D {\bf 51}, 5216 (1995), see p.~5221, hep-lat/9407017.

\bibitem{Swendsen_87} R.H. Swendsen and J.-S. Wang,
   Phys. Rev. Lett. {\bf 58}, 86 (1987).

\bibitem{Wolff_89} U. Wolff, Phys. Rev. Lett. {\bf 62}, 361 (1989).

\bibitem{Ossola-Sokal}  G. Ossola and A.D. Sokal, hep-lat/0402019 and
in preparation.

\bibitem{systematic_full} G. Ossola and A.D. Sokal, in preparation.

\bibitem{Blote_99}  H.W.J. Bl\"ote, L.N. Shchur and A.L. Talapov,
   Int. J. Mod. Phys. C {\bf 10}, 1137 (1999), cond-mat/9912005.

\bibitem{Ballesteros_99}   H.G. Ballesteros, L.A. Fern\'andez,
  V. Mart\'{\i}n-Mayor, A. Mu\~noz Sudupe, G. Parisi and J.J. Ruiz-Lorenzo,
  J. Phys. A: Math. Gen. {\bf 32}, 1 (1999), cond-mat/9805125.

\bibitem{Hasenbusch_99}  M. Hasenbusch, K. Pinn and S. Vinti,
  Phys. Rev. B {\bf 59}, 11471 (1999), hep-lat/9806012.

\bibitem{Hasenbusch_01}  M. Hasenbusch, Int. J. Mod. Phys. C {\bf 12},
   911 (2001).



% \bibitem{Knuth_68}  D.E. Knuth, {\em The Art of Computer Programming}\/,
%    vol.~2, 1st ed. (Addison-Wesley, Reading, Massachusetts, 1968).
% 
% \bibitem{Dieter_75}  U. Dieter, Math. Comp. {\bf 29}, 827 (1975).
% 
% \bibitem{Lenstra_82}  A.K. Lenstra, H.W. Lenstra, Jr. and L. Lov\'asz,
%    Math. Ann. {\bf 261}, 515 (1982).
% 
% \bibitem{Fincke_85}  U. Fincke and M. Pohst, Math. Comp. {\bf 44}, 463 (1985).
% 
% \bibitem{Graham_94}  R.L. Graham, D.E. Knuth and O. Patashnik,
%   {\em Concrete Mathematics: A Foundation for Computer Science}\/,
%   2nd ed.~(Addison-Wesley, Reading, Mass., 1994).


\bibitem{Lecuyer_99}  P. L'Ecuyer, Math. Comp. {\bf 68}, 249 (1999).

\bibitem{Salas-Sokal_00}  J. Salas and A.D. Sokal, J. Stat. Phys. {\bf 98},
   551 (2000), cond-mat/9904038.

\bibitem{Ballesteros_98}  H.G. Ballesteros and V. Mart\'{\i}n-Mayor,
   Phys. Rev. E {\bf 58}, 6787 (1998), cond-mat/9806059.


\bibitem{Ferrenberg_92}  A.M. Ferrenberg, D.P. Landau and Y.J. Wong,
   Phys. Rev. Lett. {\bf 69}, 3382 (1992).
% 
% \bibitem{Grassberger_93a}  P. Grassberger, J. Phys. A: Math. Gen. {\bf 26},
%    2769 (1993).
% 
% \bibitem{Grassberger_93b}  P. Grassberger, Phys. Lett. A {\bf 181}, 43 (1993).
% 
% \bibitem{Selke_93}  W. Selke, A.L. Talapov and L.N. Shchur,
%    JETP Lett. {\bf 58}, 665 (1993)
%    [= Pis'ma Zh. Eksp. Teor. Fiz. {\bf 58}, 684 (1993)].
% 

\bibitem{Vattulainen_94}  I. Vattulainen, T. Ala-Nissila and K. Kankaala,
   Phys. Rev. Lett. {\bf 73}, 2513 (1994), cond-mat/9406054.


\bibitem{Coddington_94}  P.D. Coddington, Int. J. Mod. Phys. C {\bf 5},
   547 (1994), cond-mat/9309017.
% 
% \bibitem{Coddington_96}  P.D. Coddington, Int. J. Mod. Phys. C {\bf 7},
%    295 (1996).
% 

% \bibitem{Sloane_on-line}
% N.J.A. Sloane, Sloane's On-Line Encyclopedia of Integer Sequences,\hfill\break
%    \verb+http://www.research.att.com/~njas/sequences/index.html+

% \bibitem{Kernighan_88}  B.W. Kernighan and D.M. Ritchie,
%    {\em The C Programming Language}\/, 2nd ed.
%    (Prentice Hall, Englewood Cliffs, N.J., 1988).
% 
% \bibitem{Metcalf_99}  M. Metcalf and J. Reid,
%    {\em Fortran 90/95 Explained}\/, 2nd ed.
%    (Oxford University Press, Oxford--New York, 1999).
% 
% \bibitem{Batrouni_85}  G.G. Batrouni, G.R. Katz, A.S. Kronfeld,
%    G.P. Lepage, B. Svetitsky and K.G. Wilson,
%    Phys. Rev. D {\bf 32}, 2736 (1985).
% 
% \bibitem{Brown_88}  F.R. Brown, N.H. Christ, Y. Deng, M. Gao and T.J. Woch,
%    Phys. Rev. Lett. {\bf 61}, 2058 (1988).

\end{thebibliography}
\end{document}